\documentclass[12pt]{article}
\usepackage{epsfig}
\usepackage{amssymb,graphicx}
\def\be{\begin{equation}}
\def\ee{\end{equation}}
\def\ba{\begin{array}{c}}
\def\ea{\end{array}}

\newcommand{\bea}{\begin{eqnarray}}
\newcommand{\eea}{\end{eqnarray}}

\begin{document}

\begin{center}

{\Large \bf {

Multi-well log-anharmonic oscillators

 }}

\vspace{13mm}

\vspace{3mm}

\begin{center}

\textbf{Miloslav Znojil}\footnote{znojil@ujf.cas.cz}
 and
\textbf{Franti\v{s}ek
R\r{u}\v{z}i\v{c}ka}\footnote{ruzicka@ujf.cas.cz}

\vspace{1mm} Nuclear Physics Institute of the CAS, Hlavn\'{\i} 130,
250 68 \v{R}e\v{z}, Czech Republic

\end{center}

\vspace{3mm}

\end{center}

\subsection*{Keywords:}

Schr\"{o}dinger equation; harmonic oscillator; logarithmic barriers;
local large$-N$ expansions;

\subsection*{PACS number:}
.

PACS 03.65.Ge - Solutions of wave equations: bound states


\subsection*{Abstract}

Large$-N$ expansions are usually applied in single-well setups. We
claim that this technique may offer an equally efficient
constructive tool for potentials with more than one deep minimum. In
an illustrative multi-well model this approach enables us to explain
the phenomenon of an abrupt relocalization of ground state caused by
a minor change of the couplings.

%
%
\newpage

\section{Introduction}

The exact solvability makes the harmonic-oscillator models of bound
states suitable for various quick estimates and qualitative
phenomenological predictions. An amended fit of observed spectra is
then usually achieved using an {\it ad hoc\,} perturbation of the
potential. In the resulting (say, one-dimensional) Schr\"{o}dinger
equation
 \be
 \left [
 -\frac{\hbar^2}{2m}
 \frac{d^2}{dx^2} +\omega^2x^2+ \lambda\,V_{(pert)}(x)
 \right ]\,\psi_n(x) = E_n\,\psi_n(x)\,,
 \ \ \ \ n = 0, 1, \ldots\,
 \label{SElog}
 \ee
the user-friendliness of the conventional Rayleigh-Schr\"{o}dinger
perturbation expansions \cite{Messiah} then explains the widespread
preference of the smooth power-law forms $V_{(pert)}(x) \sim x^{n}$
of the anharmonicities with, typically, $n=4$ or $n=6$
\cite{Fluegge}. Alas, in contrast to the wealth of the emerging
mathematical challenges \cite{Kato}, the strictly phenomenological
impact of the quartic and sextic perturbations is not too
impressive. These corrections only modify the shape of the potential
far from its minimum. As a consequence, the influence of the
perturbation is hardly felt by the low-lying, i.e., by the
experimentally most relevant, bound states.

In our recent letter \cite{Iveta2} we turned attention, therefore,
to the possible consequences of using certain less usual value of a
very small {\em negative\,} exponent $n =-\alpha \approx 0$ for
which the perturbation term $1/|x|^\alpha = 1 - \alpha\,\ln |x| +
{\cal O}(\alpha^2)$ is dominated, near the origin, by an infinitely
high but still tunnelable repulsive barrier. For the non-power-law,
short-range perturbations sampled by the logarithmic-function choice
of $V_{(pert)}(x) \sim \ln |x|$ we revealed and verified that the
study of the spiked-oscillator models of such a type may find an
unexpectedly efficient solution method in the so called large$-N$
perturbation expansions (cf., e.g., the compact review paper
\cite{Bjerrum} in this respect).

As a byproduct of the latter study we noticed that the introduction
of a ``soft'', weakly repulsive logarithmic central core in the
potential
 \be
 V_{eff}(x)=
 \omega^2x^2-g^2\,\ln |x|
 \label{soco}
 \ee
enhances the pragmatic, descriptive appeal of the model.
In the strong-coupling dynamical regime with $g^2 \gg
\omega^2$, for example, the system starts exhibiting certain
features (like a
pairwise degeneracy tendency of the low-lying even and odd states)
which are usually attributed to the double-well models with a
more strongly suppressed tunneling through the
barrier.

Due to the left-right symmetry of our illustrative double-well
soft-core model (\ref{soco}) (cf. also its square-well predecessor
in \cite{Iveta}), multiple qualitative features of the bound-state
spectra were found predictable {\it a priori}, without any extensive
{\it ad hoc\,} numerical or perturbative calculations which only
confirmed the expectations. In our paper we intend to follow and
extend this direction of research, therefore. We will consider
certain more complicated shapes of the potentials with more than one
logarithmic repulsive spike.


\section{Large$-N$ method {\it in nuce}\label{themethod}}

The most elementary toy-model two-particle interactions $V(x)$ used
in atomic, molecular and nuclear physics are very often composed of
an asymptotically attractive harmonic-oscillator potential
$V_{(attractive)}(x)=\omega^2x^2$ and of its short-range repulsive
component. Typically, $V_{(repulsive)}(x)=g^2/x^2$ is used, at any
number of particles $A$, in the popular $A-$body Calogero model
\cite{Calogero}. This model is exactly solvable and, hence,
suitable for our introductory illustrative and
methodical purposes.

\subsection{Calogero model in the strong-repulsion dynamical regime\label{gIa}}

In units such that $\hbar=2m=1$ the exactly solvable
one-dimensional Schr\"{o}dinger equation
 \be
 \left [
 -\frac{d^2}{dx^2} + \frac{g^2}{x^2}+\omega^2x^2
 \right ]\,\psi_n(x) = E_n\,\psi_n(x)\,,
 \ \ \ \ n = 0, 1, \ldots\,
 \label{SErad}
 \ee
with an impenetrable central barrier can be interpreted not only as
the $A=2$ special case of the one-dimensional $A-$particle Calogero
model (in which the attractive harmonic-oscillator two-body force is
complemented by strong repulsion at short distances \cite{Calogero})
but also as a conventional radial component of a $D-$dimensional
harmonic oscillator \cite{Fluegge}. In both of these contexts,
unfortunately, the barrier is impenetrable, not admitting a
tunneling. As a consequence, not only the most common radial
Schr\"{o}dinger equation but also the more sophisticated Calogero's
equation must be perceived as living on a half-line or, in the
Calogero's case, as describing two independent systems defined in
two separate ``Weyl chambers'' with $x \in (-\infty,0)$ and $x \in
(0,\infty)$, respectively.

Any spontaneous transfer of the state of the system to the other
chamber is, in the model, excluded. One must conclude that from the
point of view of phenomenology the main weak point of the latter
centrifugal-type repulsion is that it is impenetrable. Even in the
most elementary two-body Calogero model the apparently double-well
dynamics must be interpreted, in physics, as a pair of two
independent single-well problems. At the same time, the model with
the special choice of $g^2 \gg \omega^2$  may serve as an
illustrative example of the above-mentioned large$-N$ expansion
techniques.

%
%
%
%

\subsection{Oscillations near the deep local minima}

The very essence of the efficient perturbative strong-coupling
large$-N$ expansion technique lies in the approximation of the
interaction. Near the deep minimum of $V_{eff}(x)$ at $x=R=R(N)$ the
interaction is approximated by its {\em truncated\,} Taylor series,
 \be
 V_{eff}(x)\approx
  c_0+ c_1(x-R)+ c_2(x-R)^2
 +c_3(x-R)^3+\ldots
 +c_M(x-R)^M\,,
 \ \ \ \ \ c_1=0
 \,.
 \label{potapro}
 \ee
One of the most persuasive illustrations of the amazing practical
numerical efficiency of such an approach is provided by the radial
harmonic-oscillator Schr\"{o}dinger equation
 \be
 \left(-\frac{d^2}{dx^2} + x^2+\frac{N(N+1)}{x^2}
 \right)\ \psi_m(x) = E_m\ \psi_m(x)\,,
 \ \ \ \ \ N \gg 1
\,,\ \ \ \  m = 0, 1, \ldots\,
 \label{equati}
 \ee
in which the minimum of the complete effective interaction
$V_{eff}(x)=x^2+{N(N+1)}/{x^2}$ lies at $x=R(N)=[N\,(N+1)]^{1/4}\gg
1$. In the Taylor series (\ref{potapro}) we easily evaluate
$c_0=V_{eff}(R) \gg 1$, $c_2 = {\cal O}(1)$ and $c_{2+j} = {\cal
O}(1/R^{j})$, $j=1,2,\ldots$. Thus, via an {\it ad hoc\,} shift of
coordinate $x \to x-R$ in Eq.~(\ref{SElog}) we found the constant
$\omega^2=c_2$ as well as the desired small parameter
$\lambda=1/R(N)$.

Needless to add that even the first nontrivial truncations of
expansion (\ref{potapro}) yield already a fairly reliable
perturbative low-lying spectrum via Eq.~(\ref{SElog}) (cf.
\cite{Iveta2}). The problems only arise when we imagine that the
approximate wave functions lie, by construction, on the whole real
line, $x \in (-\infty,\infty)$. This appears to be a decisive
conceptual shortcoming of the application of the method to
Eq.~(\ref{equati}) because as long as the centrifugal barrier does
not admit tunneling, the exact bound states $\psi_m(x)$ do only live
on the half-axis, $x \in (0,\infty)$. For this reason, the large$-N$
approximants cannot converge \cite{Bjerrum,Arnold}.

In practice, fortunately, the $M \to \infty$ divergence of the
Taylor-series potentials (\ref{potapro}) is only rather weakly felt
by the low lying approximate energies themselves. In the related
literature, numerous tests of their $M-$th order large$-N$ {\it
alias\,} small$-\lambda$ representation
 \be
 E_m(\lambda) = E_m(0) + \lambda\,E_m^{(1)} + \lambda^2\,E_m^{(2)} +
 \ldots + \lambda^M\,E_m^{(M)} + {\cal O}(\lambda^{M+1})\,
 \label{truseries}
 \ee
were performed for many phenomenological single-well-dominated
interactions $V_{eff}(x)$ \cite{laN}. Most of these tests confirmed
that the loss of the precision (reflecting the $M\to \infty$
divergence) only starts to influence the reliability of the results
at certain optimal perturbation-expansion orders
$M=M_{optimal}(\lambda)<\infty$.

\subsection{Penetrable barriers}

\subsubsection{Complexified centrifugal term}

One of the first amendments of the chamber-separation arrangement
has been found in the framework of ${\cal PT}-$symmetric quantum
mechanics. In this formulation of quantum theory
\cite{Carl,ali,lotor} the tunneling between Weyl chambers implying a
``relocalization'' of the system on the real line of $x$ has been
rendered possible via an {\em ad hoc\,} regularization of the
centrifugal-like barrier by its complexification,
 \be
 \frac{g^2}{x^2} \ \to \ \frac{g^2}{(x-{\rm i}\varepsilon)^2}
 = \frac{x^2-\varepsilon^2}{(x^2+\varepsilon^2)^2}
 + {\rm i} \frac{2\,\varepsilon\,x\ }{(x^2+\varepsilon^2)^2}\ .
 \label{ptbar}
 \ee
At $A=2$ \cite{ptho} and at $A=3$ \cite{Tater} it has been shown
that in spite of the manifest non-Hermiticity of the complexified
Calogero Hamiltonians the spectrum of the energies remains real and
given in closed form. Reflecting, nicely, the nontrivial effects and
consequences of the tunneling.

The latter two proposals using non-Hermitian interactions remained
incomplete because the construction of the related physical Hilbert
space (i.e., of a nontrivial inner-product metric $\Theta$ yielding
the correct probabilistic interpretation of wave functions) proved
prohibitively difficult \cite{Chen}. In the light of some recent
rigorous mathematical analyses, moreover, open questions still
concern even the very existence of {\em any\,} inner-product metric
in such a local-interaction class of non-self-adjoint Hamiltonians
\cite{scatt,ATbook}.

\subsubsection{Central logarithmic spike}

An easier way towards an amended model with tunneling has
subsequently been proposed in \cite{Tater2}. In the framework of an
entirely conventional quantum mechanics we merely replaced the
complex spike (\ref{ptbar}) by the most elementary Hermitian
point-interaction delta-function barrier which admits tunneling as
well. We concluded that the property of the impenetrability of the
centrifugal barrier in Eq.~(\ref{equati}) is, for the reliability of
the method, inessential.

The analytic, smooth logarithmic barrier of Refs.
\cite{Iveta2,Iveta} emerges as one of the other eligible candidates
for a partially penetrable barrier, therefore. In {\it loc. cit.} we
replaced the exactly solvable radial bound-state problem
(\ref{SErad}) by the harmonic oscillator Schr\"{o}dinger equation
perturbed by the logarithmic repulsive spike,
 \be
 \frac{g^2}{x^2} \ \to \ g^2\,\ln\,\frac{1{\ }}{x^2} =
 -2\,g^2\,\ln\,|x|\,.
 \label{replac}
 \ee
The new Schr\"{o}dinger equation has been found solvable by the
large$-N$ perturbation expansion technique which proved applicable
at all of the sufficiently large couplings $g^2 \gg \omega^2$. One
of the encouraging technical merits of the replacement
(\ref{replac}) of the power of $x$ by the more complicated
logarithmic function has been found in the not quite expected
user-friendliness of algebraic manipulations. This was a discovery
which served also as an initial inspiration of our present paper.

\section{Potentials with multiple logarithmic spikes}


\begin{figure}[h]                    
\begin{center}                         
\epsfig{file=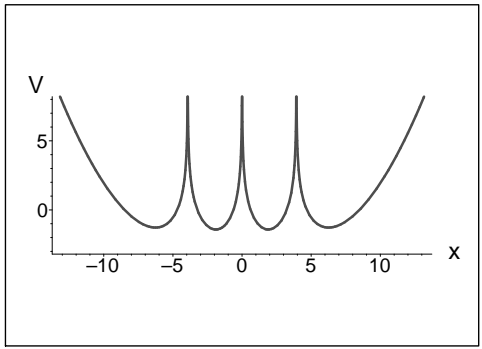,angle=270,width=0.36\textwidth}
\end{center}    
\vspace{2mm} \caption{Illustrative four-well potential
$V(x)=0.11\,x^2-\ln x^2  -\ln (0.11\,x^2-1.7)^2$.
 \label{trija}
 }
\end{figure}

 \noindent
In the present paper we will pay attention to the following
generalization of Eq.~(\ref{replac}),
 \be
  g^2\,\ln \frac{1}{x^2} \ \to \  g^2\,\ln \frac{1}{x^2}+
 \sum_{j=1}^{K}\,h^2_j\,
 \left (
 \ln \frac{1}{(x-s_j)^2}+
 \ln \frac{1}{(x+s_j)^2}
 \right )
  = -g^2\,\ln x^2 -
 \sum_{j=1}^{K}\,h^2_j\,\ln (x^2-s_j^2)^2\,.
 \ee
Parameters $s_j>0$, $j=1,2,\ldots,K$ represent the (left-right
symmetric) positions of the unbounded but still transparent
logarithmic spikes converting the harmonic-oscillator well into a
multi-well potential. An illustrative sample of its shape is
displayed in Fig.~\ref{trija}. In a way inspired by such a
quadruple-well example we decided to make the large$-N$ approach
``localized'', restricted to any separate (and, presumably, still
sufficiently pronounced and deep) minimum of the potential. We
imagined that such an approach could open the way, e.g., towards a
better understanding of the role of the small changes of the
parameters which might influence, in a not entirely trivial manner,
not only the positions of the separate minima of $V(x)$ but also the
widths of the wells near these minima.

In the low-lying part of the spectrum such a simplification of the
problem may be expected to enhance our understanding of what happens
with the role of the minima of the potential which represent the
eligible stable equilibria in classical systems. Naturally, this
possibility follows from the absence of the tunneling so that the
related picture of dynamics cannot be easily transferred to the
quantum models with tunneling.

\subsection{Large$-N$ pattern at $K=0$}

After the ``softening'' (\ref{replac}) of the barrier the spatial
symmetry of the $K=0$ effective potential with $g \neq 0$ (and, say,
with
 $x \in
(0,\infty)$ in Eq.~(\ref{soco})) still enables us to deduce that
 \be
 V(x)=\omega^2x^2-g^2 \ln x^2\,,\ \ \ \ \ \
 V'(x)=2\,\omega^2 x-2g^2/ x\,,\ \ \ \ \ \
 V''(x)=2\,\omega^2+2g^2/ x^2\,,\ \ldots\,.
 \label{billirub}
 \ee
This localizes the minimum of the potential at $x=x_{min}=R=g/\omega
$ and defines the small parameter $\lambda=1/R$. With
$e=2.718\ldots$ and $V'(R)=0$ we have
 \be
 V(R)=\omega^2R^2-2g^2 \ln R=g^2\,
 \left (1-2 \ln g+2 \ln \omega
 \right )=g^2\,\ln (e\,\omega^2/g^2)
 \,,\ \ \ \ \ \
 V''(R)=4\,\omega^2\,,\ \ldots\,
 \label{extrab}
 \ee
in the truncated Taylor series (\ref{potapro}). One also easily
defines the depth and width of the leading-order harmonic-oscillator
potential and arrives at the leading-order energies as prescribed by
Eq.~(\ref{truseries}),
 $$
 E_n=g^2\,\ln (e\,\omega^2/g^2)+\sqrt{2}(2n+1)\,\omega +
 {\cal O}(1/g)
 \,\,,\ \ \ n =
 0, 1, \ldots\,.
 $$
For the sufficiently strong repulsion $g \gg 1$, these values are
found to compare well with the brute-force numerical results (cf.
\cite{Iveta2}).

\subsection{Large$-N$ pattern at $K>0$}

For our present purposes, one of the key consequences of the mere
marginal relevance of the convergence or divergence of the
asymptotic  ``large$-N$ input'' Taylor series (\ref{potapro}) is
that the reliability of the perturbation approximants as provided by
expansions (\ref{truseries}) is almost exclusively dependent on the
{\em local\,} depth and width of the potential well near its
minimum. In other words, without any real loss of the reliability of
the results one can admit the existence of arbitrarily many other,
separate minima. Such a conclusion leads us to the very core of our
present methodical message: Under the assumption of a sufficiently
suppressive separation barriers between the neighboring minima of
the potential, one can apply the large$-N$ approximation technology,
separately, in every individual well. Subsequently, the key idea is
that the global low-lying spectrum will be mostly localized in the
deepest and widest individual well. A delocalization of the system
may be expected to occur only in the ``exceptional'' scenarios in
which there will be no clearly dominant single individual well (or,
in the spatially symmetric arrangements, a symmetric non-central
pair of dominant wells).

\begin{figure}[h]                    
\begin{center}                         
\epsfig{file=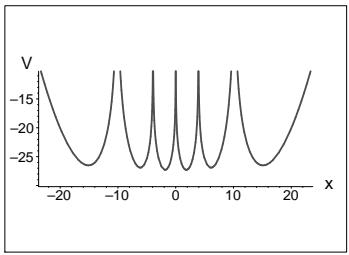,angle=270,width=0.26\textwidth}
\end{center}    
\vspace{2mm} \caption{The shape of $V(x)=\omega^2\,x^2-g^2\,\log
x^2-\lambda^2\,\log (\omega^2\,x^2-h^2)^2
 -\mu^2\,\log (\omega^2\,x^2-f^2)^2$
at $\omega^2=0.11,\,g^2=2\,,\lambda^2=2.3\,,h^2=1.7,\, \mu^2=5$ and
$f^2=11.3$.
 \label{tripja}
 }
\end{figure}

The presence of {\em several\,} repulsive logarithmic spikes in the
potentials will form a multiple-well potential with its minima
separated by the logarithmic barriers which are unbounded but still
penetrable. Due to the freedom in our choice of the number of the
barriers as well as of their strengths and positions, a remarkable
flexibility of the resulting shape of the potentials will be
achieved. Fig.~\ref{trija} offers a typical illustration in which
the quadruple-well shape of the potential is specific in having also
the comparable depths of the separate minima. An analogous
multi-well shape of our potentials may be also obtained at $K=2$
(cf. Fig.~\ref{tripja}), etc.

Remarkably enough, the latter descriptive merit of the model proves
accompanied by an enhanced sensitivity of the depth of the separate
minima to the comparatively small changes of the parameters. In the
context of classical physics such a sensitivity is usually
interpreted as opening a way towards a ``catastrophe'' \cite{Thom}.
In our present quantum-physics setting, such a sensitivity to the
parameters must be interpreted more carefully of course
\cite{catast}.

\subsection{An interplay between the widths and depths}


In a way encouraged by the user-friendliness of the single-spike
logarithmic $K=0$ anharmonic oscillator (cf. Eq.~(\ref{billirub}))
and of its large$-N$ description we believe that the study of the
generalized, multi-well and left-right symmetric potential
 \be
 V_{(K)}(x) = \omega^2x^2+
 g^2\,\ln \frac{1}{x^2}+
 \sum_{j=1}^{K}\,h^2_j\,
  \ln \frac{1}{(x^2-s_j^2)^2}\,
  \label{intera}
 \ee
(sampled in Fig.~\ref{tripja} at $K=2$) might be rewarding due to
the variability of the integer $K=1, 2, \ldots\,$. At the not too
large values of $K$ the purpose of the approximate determination of
the spectra may be seen in the prediction of the points of the
relocalization instabilities. Indeed, the comparison of the
large$-N$ spectra in the individual wells is probably one of the
most reliable tools of the determination of the exceptional,
``critical'', degeneracy-simulating sets of the parameters.

Any path defined in the space of
parameters and passing through such a ``relocalization'' instability
may be interpreted as an instant of transition of the system
from its initial ground state localized, say, near the origin, to
another ground state with probability density
which is, near the origin, strongly suppressed.
One can expect the experimental observability of such a phenomenon,
especially in the low-lying-states setup.

In the context of theory we are, naturally, free to prefer working
in the dynamical regime with strong repulsion in which the
user-friendly large$-N$ solutions will lie sufficiently close to the
exact ones. In our present paper we feel guided by the observation
that a successful description of a given quantum system is often
facilitated by the occurrence of a small parameter in the
Hamiltonian, $H=H(\lambda)$. Typically, such a knowledge leads to a
more or less routine expansion of the observable quantities in the
powers of $\lambda$. In the multi-well potentials with a partially
suppressed tunneling such an approach is simply to be implemented
{\em locally}.

In the context of mathematics we intend to re-emphasize that the
power series of perturbation theory need not be required convergent.
The usefulness of the divergent {\it alias\,} asymptotic series is
best sampled by the large$-N$ expansion technique. The specific
features of the technique enable one to use it for the description
of qualitative aspects of the multiple-well dynamical scenarios.


\section{Approximate bound states in individual wells
\label{atheposi}}

The generalized quantum bound-state problem
 \be
 \left [
 -\frac{d^2}{dx^2} + V_{(K)}(x)
 \right ]\,\psi_n(x) = E_n\,\psi_n(x)\,,
 \ \ \ \ n = 0, 1, \ldots\,,
 \ \ \ \ \psi_n(x) \in L^2(\mathbb{R})\,
 \label{SEmultumesc}
 \ee
is not too easily solvable even by the dedicated numerical methods.
For this reason, its large$-N$ tractability would be welcome. In the
light of our preceding comments, the approximate evaluation of the
low-lying bound states may be expected helpful, especially when one
of the wells dominates by its deptsh and width, and especially when
the coupling constants $g^2$ and $h_j^2$ remain {\em all\,}
sufficiently large. Under such a restriction the various,
topologically different multi-well versions  Schr\"{o}dinger
bound-state problem (\ref{intera}) + (\ref{SEmultumesc}) for
low-lying states may still be given a user-friendly, perturbatively
solvable form (\ref{SElog}).

\subsection{Truncated Taylor series}

Once we decided to study Eq.~(\ref{SEmultumesc}) with potentials
(\ref{intera}) in the strongly-spiked interaction regime, we feel
entitled to split the wave functions into their separate (and, at
the end, mutually matched)) components restricted just to one of the
(presumably, deep) wells. This simplifies the general $2K-$ or
$(2K+1)-$ barrier dynamical scenario and enables us to try to
describe the bound states in an approximate, semi-qualitative
manner.

The success of the strategy depends on several factors including the
specification of the potential and the sensitivity of its shape to
the variations of the parameters. The task is facilitated by several
formal merits of our choice of the form of the potential. The key
merit concerns the construction of the Taylor series for which one
needs to know the derivatives of the potential. It is immediate to
verify that the latter evaluation is straightforward, mainly due to
the disappearance of the complicated logarithmic function after the
differentiation. Thus, the elementary manipulations yield the
necessary formulae
 $$
 V'_{(K)}(x) = 2\,\omega^2x -\frac{2\,g^2}{x} -2\,
 \sum_{j=1}^{K}\,h^2_j\,\frac{2x}{x^2-s_j^2}\,,
 $$
 $$
 V''_{(K)}(x) = 2\,\omega^2 +\frac{2\,g^2}{x^2} +4\,
 \sum_{j=1}^{K}\,h^2_j\,\frac{x^2+s_j^2}{(x^2-s_j^2)^2}
 $$
and
 $$
 V'''_{(K)}(x) = -\frac{4\,g^2}{x^3} -8x\,
 \sum_{j=1}^{K}\,h^2_j\,\frac{x^2+3\,s_j^2}{(x^2-s_j^2)^3}
 $$
etc. In their light, the classification of the possible dynamical
scenarios degenerates to the comparatively straightforward algebraic
manipulations.


\subsection{Example: Central well at $g=0$\label{eposi}}

At $g=0$ and at any $K$, one of the minima of  $V_{(K)}(x)$
lies in the origin, $x=x_0=0$. At $K>0$ the
locally minimal
value of the potential is negative,
 $$
 V_{(K)}(0) =
 -4\, \sum_{j=1}^{K}\,h^2_j\,
  \log {s_j}
 $$
(one should add that $V_{(0)}(0)=0$ at $K=0$). The local $M=2$
Taylor-series approximation of the potential degenerates to harmonic
oscillator,
 $$
 V_{(K)}(x)=V_{(K)}(0) + \frac{1}{2}\,V''_{(K)}(0)x^2+{\cal O}(x^3)
 \approx V_{(K)}(0) + \Omega_{(K)}^2\,x^2\,.
 $$
Explicit formula is available for the real and positive
 $$
 \Omega_{(K)}=\sqrt{\omega^2  +
 \sum_{j=1}^{K}\,\frac{2\,h^2_j}{s_j^2}}\ .
 $$
Thus, whenever the central minimum is a global minimum, we may
deduce the leading-order formula for the low-lying energies
 \be
 E_n=V_{(K)}(0)+(2n+1)\Omega_{(K)}+ {\rm corrections}\,,
 \ \ \ \ n = 0, 1, \ldots \,.
 \ee
Otherwise, this formula just represents such a subset of the bound
states for which the probability density is concentrated near the
origin.

\subsection{The role of parameters at $K=1$}

\subsubsection{Triple-well model with $g=0$}

Besides the above-described central well the models with $g=0$ may
exhibit also a $K-$plet of double-well non-central minima. In
particular, in the first nontrivial case with $K=1$ the potential
has the two off-central minima at $x=R_{(\pm)}$ where, by
definition,
 $$
 R_{(\pm)}=\pm \sqrt{s^2+\frac{2h^2}{\omega^2}}\,.
 $$
The two identical values of the other two local (and equal) extremes
of the potential
 $$
 V_{}(R_{(\pm)})=\omega^2s^2+2h^2 +h^2\log  \frac{\omega^4}{4h^4}\,
 $$
lie at the other two non-central minima. Due to the positivity of
the second Taylor-series coefficient
 $$
 c_2= \Omega_{(\pm)}^2={\frac{1}{2} V''(R_{(\pm)}) }=
 {\omega^2 +2\,
 h^2\,\frac{R^2+s^2}{(R^2-s^2)^2}}
 ={2\omega^2 +\frac{s^2\omega^4}{h^2}}\,
 $$
we may evaluate the almost degenerate pair of the approximate
double-well energies
 \be
 E_n^{(\pm)}=
 \omega^2s^2+2h^2 +h^2\log  \frac{\omega^4}{4h^4}
 +(2n+1)\omega\,
 \sqrt{2 +\frac{s^2\omega^2}{h^2}}
 + {\cal O}(1/h)\,,
 \ \ \ \ n = 0, 1, \ldots \,.
 \ee
This formula complements the central harmonic-oscillator
approximation of paragraph \ref{eposi}. In the generic case one of
these parts of the spectrum is dominant (i.e., low-lying) while the
other one remains highly excited.

At an exceptional instant of the relocalization catastrophe both of
these candidates for the ground state (as well as, in principle, for
the first few low lying excited states) remain comparable. In this
case the leading-order large$-N$ approximation ceases to be
applicable. The exact numerical solutions must be constructed
instead.

\subsubsection{Quadruple-well model with $g\gg \omega$}

At $K=1$ the condition $V'_{(K)}(R) = 0$ for an extreme at
$x=R=\sqrt{Z} \neq 0$ has the form
 $$
  \omega^2Z -{g^2} -
 h^2\,\frac{2Z}{Z-s^2}=0\,.
 $$
This is a quadratic equation yielding the two positive roots
 $$
 Z_\pm =\frac{1}{2}(a+c\pm \sqrt{ 2\,a\,c+b^2})\,,
 \ \ \ \
 a=s^2+g^2/\omega^2\,,
 \ \ \ \
 b=s^2-g^2/\omega^2\,,
 \ \ \ \
 c=2\,h^2/\omega^2\,.
 $$
Their insertion leads immediately to a lengthy but explicit
algebraic formula for the value of the coefficient $c_2=1/2\,
V''_{(1)}(\sqrt{Z_\pm})$ entering the Taylor series (\ref{potapro})
which defines the approximate harmonic-oscillator potential.
Subsequently, one immediately obtains the low-lying energy spectrum
for the states which are localized near the respective local minimum
of the global potential function  $V_{(1)}(x)$.

%
%

\section{Conclusions\label{555}}

In the generic multi-well systems the descriptions of bound states can
rarely be performed by non-numerical means. In our paper we verified
that a slightly modified version of the large$-N$ approximation
approach may simplify the constructions and offer an alternative,
more straightforward
mathematical tool.

In the context of quantum physics we paid
particular attention to the fact that the control of the coupling
constants in the potential immediately controls also the occurrence,
properties and localization of the low-lying bound states. We
emphasized that in the vicinity of a certain exceptional set of
parameters, a comparatively small change of these parameters may
lead to an abrupt ``relocalization'' jump in the particle
probability density.

The presence of the repulsive
barriers has been shown to play a decisive role in the possible
interpretation of
the states which are highly sensitive to the changes of the external
conditions. Such states can be perceived as quantum
analogues of the classical systems passing through an instability.
Naturally, the analogy is incomplete, mainly because the
Thom's classification of the classical ``catastrophes'' did not find
its sufficiently universal
quantum-theoretical counterpart in
mathematical literature yet \cite{catast}.

\subsection{Double-well models and quantum catastrophes}
%
%
%
%

One of the most characteristic features of {\em any\,} bound-state
Schr\"{o}dinger equation (\ref{SEmultumesc})
%
with {\em any\,} conventional symmetric double-well interaction
potential $V(x)=V(-x)$ is that the first excited-state energy $E_1$
does not lie too far from its ground-state predecessor $E_0$.
Intuitively, the phenomenon is explained by the existence of a
central repulsive barrier which suppresses the central part of the
wave function. The
energy of the even ground-state wave function $\psi_0(x)$
without a nodal zero lies close to its
first-excitation
partner
and odd wave function
$\psi_1(x)$ possessing a single nodal zero in the origin.

In Ref.~\cite{Iveta2} we have shown that such a level-degeneracy
tendency is observed, in the lowest part of the spectrum at least,
even for the very weak (viz., logarithmic) central repulsive
barriers such that $V(x) \sim \ln (1/x^2) + {\cal O}(1)$ near the
origin. At the same time, the effect may become quickly lost after
the breakdown of the spatial symmetry of the potential. In general,
one of the minima then becomes perceivably deeper and starts playing
the dominant role in the localization of the low-lying wave
functions. The closest classical analogue of such a phenomenon can
be seen in the Thom's catastrophes called ``fold'' or ``cusp''
\cite{Zeeman}.

\subsection{More wells}

In the context of the classical catastrophe theory \cite{Zeeman}
it is rather surprising to notice that in the literature,
not too much attention is being paid to the more general
quantum dynamical
scenarios in which the number of the ``tunable'' minima of potential
$V(x)$ is chosen greater than two.
In our present paper we outlined
a way towards the
simulations of the
less elementary quantum catastrophes.
The tests of the idea were performed using a model with several
application-oriented merits. One of them may be seen in the form of
the individual barriers which remain singular (i.e., unbounded) but
still much weaker than any power of $x$. Hence, the barriers admit a
tunneling which is comparatively intensive even in the not too high
energy levels. Still, the existence of the barriers leads to the
localization of the wave functions near the deep minima of the
potential.
One only has to
keep in mind that the lowest, ground state is often
localized in the widest rather than in the deepest well or wells.

A
subtler
interplay between the separate individual
wells only enters the game when none of the approximate
ground-state energy-level candidates happens to dominate.
This implies that
the wave functions become ``delocalized'', spread over several
competing wells.
In this critical dynamical regime
our present method based on the identification of
the dominant well (or rather of the dominant pair of wells) ceases
to be applicable.
This being said, even the use of approximants
enables us to study the forms and alternative scenarios of the
unfolding of the eligible
quantum-catastrophic phenomena.

\subsection*{Acknowledgements}

The project was supported by GA\v{C}R Grant Nr. 16-22945S.


\end{document}